\documentclass[twocolumn,aps,prl,superscriptaddress,footinbib,amsmath,amssymb,aps,floatfix]{revtex4-1}

\usepackage{bm,graphicx}

\newcommand{\up}{|\!+\!z\rangle}
\newcommand{\down}{|\!-\!z\rangle}
\newcommand{\updown}{|\!\pm z\rangle}
\newcommand{\probe}{\mid\!\!\mathrm{pr}\rangle}
\newcommand{\TTFi}{(T/T_{\mathrm{F}})_\mathrm{i}}
\newcommand{\taur}{\tau_{\mathrm{r}}}
\newcommand{\svec}[1]{\bm{#1}}

\newcommand{\grad}{\nabla_{\!j}}
\newcommand{\T}{{\mathcal{T}}} 
\newcommand{\lnkfa}{{\ln(k_\mathrm{Fi} a_\mathrm{2D})}}
\newcommand{\kff}{{k_\mathrm{Ff} }}
\newcommand{\kfi}{{k_\mathrm{Fi} }}
\newcommand{\Contact}{{\mathcal{C}_\mathrm{2D} }}

\begin{document}

\title{Observation of quantum-limited spin transport in strongly interacting two-dimensional Fermi gases}
\author{C.\ Luciuk}
\author{S.\ Smale}
\affiliation{Department of Physics, University of Toronto, Ontario M5S\,1A7 Canada}
\author{F.\ B\"ottcher}
\affiliation{5.\,Physikalisches Institut and Centre for Integrated Quantum Science and Technology, Universit\"at Stuttgart, D-70569 Stuttgart, Germany}
\author{H.\ Sharum}
\author{B.\ A.\ Olsen}
\author{S.\ Trotzky}
\affiliation{Department of Physics, University of Toronto, Ontario M5S\,1A7 Canada}
\author{T.\ Enss}
\affiliation{Institut f\"ur Theoretische Physik, Universit\"at Heidelberg, D-69120 Heidelberg, Germany}
\author{J.\ H.\ Thywissen}
\affiliation{Department of Physics, University of Toronto, Ontario M5S\,1A7 Canada}
\affiliation{Canadian Institute for Advanced Research, Toronto, Ontario M5G\,1Z8 Canada}

\date{{\today}}
\begin{abstract}
We measure the transport properties of two-dimensional ultracold Fermi gases during transverse demagnetization in a magnetic field gradient. Using a phase-coherent spin-echo sequence, we are able to distinguish bare spin diffusion from the Leggett-Rice effect, in which demagnetization is slowed by the precession of spin current around the local magnetization. When the two-dimensional scattering length is tuned to be comparable to the inverse Fermi wave vector $k_F^{-1}$, we find that the bare transverse spin diffusivity reaches a minimum of $1.7(6)\hbar/m$, where $m$ is the bare particle mass. The rate of demagnetization is also reflected in the growth rate of the $s$-wave contact, observed using time-resolved spectroscopy. At unitarity, the contact rises to $0.28(3) k_F^2$ per particle, measuring the breaking of scaling symmetry. Our observations support the conjecture that in systems with strong scattering, the local relaxation rate is bounded from above by $k_B T/\hbar$.
\end{abstract}

\maketitle

Conjectured quantum bounds on transport appear to be respected and nearly saturated by quark-gluon plasmas \cite{Kovtun:2005tz,*Schafer:2009vf,*Adams:2012ki,SONG2013114c}, unitary Fermi gases \cite{Thomas2010,*Elliott:2014bv,*Joseph:2015kt,Enss:2011,*Enss12,Wlaziowski:2012ei,*Wlaziowski:2015ek,Zwierlein2011,*Sommer:2011ig,*Valtolina:2016,Bardon:2014, Trotzky:2015fe,Bruun:2011ue,*Bruun:2011hn,Levin:2011,*Duine:2012,*Heiselberg:2012vd,*Huse:2012,*Chevy:2012,*Enss:2012b,Enss:2013ti}, and bad metals \cite{Bruin:2013hc,*Zhang:2016uz,Hartnoll:2015kj}. For many modalities of transport these bounds can be recast as an upper bound on the rate of local relaxation to equilibrium $1/\taur \lesssim k_B T/\hbar$, where $k_B$ is the Boltzmann constant and $T$ is  temperature \cite{Sachdev,Zaanen:2004he}. Systems that saturate this ``Planckian'' bound do not have well defined quasiparticles promoting transport \cite{Sachdev,Zaanen:2004he,Kovtun:2005tz,*Schafer:2009vf,*Adams:2012ki,Bruin:2013hc,*Zhang:2016uz,Hartnoll:2015kj}. A canonical example is the quantum critical regime, where one expects diffusivity $D \sim \hbar/m$, a ratio of shear viscosity to entropy density $\eta/s \sim \hbar/k_B$, and a conductivity that is linear in $T$ \cite{Enss:2011,*Enss12,Bruin:2013hc,*Zhang:2016uz,Hartnoll:2015kj}. These limiting behaviors can be understood by combining $\taur$ with a propagation speed $v \sim \sqrt{k_B T/m}$, for example $D\sim v^2 \taur$. This argument applies to ultracold three-dimensional (3D) Fermi gases, whose behavior in the strongly interacting regime is controlled by the quantum critical point at divergent scattering length, zero temperature, and zero density \cite{Nikolic07,Veillette07,Enss:2011,*Enss12}. In such systems, one observes $D \gtrsim 2 \hbar/m$ \cite{Zwierlein2011,*Sommer:2011ig,*Valtolina:2016, Bardon:2014, Trotzky:2015fe} and $\eta/s \gtrsim 0.4 \hbar/k_B$ \cite{Thomas2010,*Elliott:2014bv,*Joseph:2015kt}, compatible with conjectured quantum bounds.

However in attractive two-dimensional (2D) Fermi gases, scale invariance is broken by the finite bound-state pair size, so the strongly interacting regime is no longer controlled by a quantum critical point \cite{Nikolic07,Hofmann:2011ds,Bruun:2012,*Enss:2012a,Schafer:2012,Hofmann:2012ic,Taylor:2012if,Levinsen:2015dt}. Strikingly, an extreme violation of the conjectured $D \gtrsim \hbar/m$ bound has been observed in an ultracold 2D Fermi gas: a spin diffusivity of $6.3(8) \times 10^{-3} \hbar/m$ near $\ln(k_{\rm{F}} a_\mathrm{2D})=0$ \cite{Kohl2013}, where $k_{\rm{F}}$ is the Fermi momentum and $a_\mathrm{2D}$ is the 2D $s$-wave scattering length. No similarly dramatic effect of dimensionality is observed in charge conductivity \cite{Bruin:2013hc,*Zhang:2016uz} or bulk viscosity \cite{Vogt:2012}, and such a low spin diffusivity is unexplained by theory \cite{Bruun:2012,*Enss:2012a,Enss:2013ti}.

In this work, we recreate the conditions of Ref.~\cite{Kohl2013}, and study the demagnetization dynamics of ultracold 2D Fermi gases using both a coherent spin-echo sequence \cite{Trotzky:2015fe} and time-resolved spectroscopy \cite{Bardon:2014}. We find a modification of the apparent diffusivity by the Leggett-Rice (LR) effect \cite{LR:1968,*Leggett:1970}, however, in disagreement with Ref.~\cite{Kohl2013}, we find that the quantum bound for the spin diffusivity is satisfied in all conditions accessible to our apparatus. Near $\ln(k_{\rm{F}} a_\mathrm{2D})=0$, where the minimum diffusivity is observed, we quantify the breaking of scale invariance by measuring the contact, whose magnitude suggests that the gas is in a many-body excited state during demagnetization.

Our experiments use the three lowest-energy internal states, labeled $\down$, $\up$, and $\probe$, of neutral $^{40}$K atoms. Interactions between $\down$ and $\up$ atoms are tuned by the $s$-wave Feshbach resonance \cite{RevModPhys2010} at $202.1$\,G, while $\probe$ atoms remain weakly interacting with $\updown$ atoms, and any atoms in identical spin states are non-interacting since the gas is ultracold. An ensemble of 2D systems is prepared by loading a 3D cloud of $\down$ atoms into an optical lattice with a period of 380\,nm along the $x_3$ direction \cite{SM}. At the final lattice depth of  $V_0=50$\,$E_R$, where $E_R/\hbar \simeq 2 \pi \times 8.64\,$kHz, the 2D samples are isolated from one another and in near-harmonic confinement with $\omega_3 \simeq 2\pi \times 122\,$kHz. The transverse confinement with  $\omega_{1,2} \simeq 2\pi \times 600\,$Hz is controlled by an optical dipole trap. The reduced temperature $\TTFi$ of the 2D ensemble can be varied between 0.20 and 1.20, where $T_{\rm{F}} \equiv E_{\rm{F}}/k_B$, and $E_{\rm{Fi}}=\hbar^2 k_{\rm{Fi}}^2/2m$ is the Fermi energy of the central 2D system in its initial polarized state. We parametrize the interaction strength by $\lnkfa$. A static magnetic field gradient $B'$ along $x_1$ is set to $20.3(2)$\,G/cm unless stated otherwise.

\begin{figure}[tb!]
\includegraphics[width=\columnwidth]{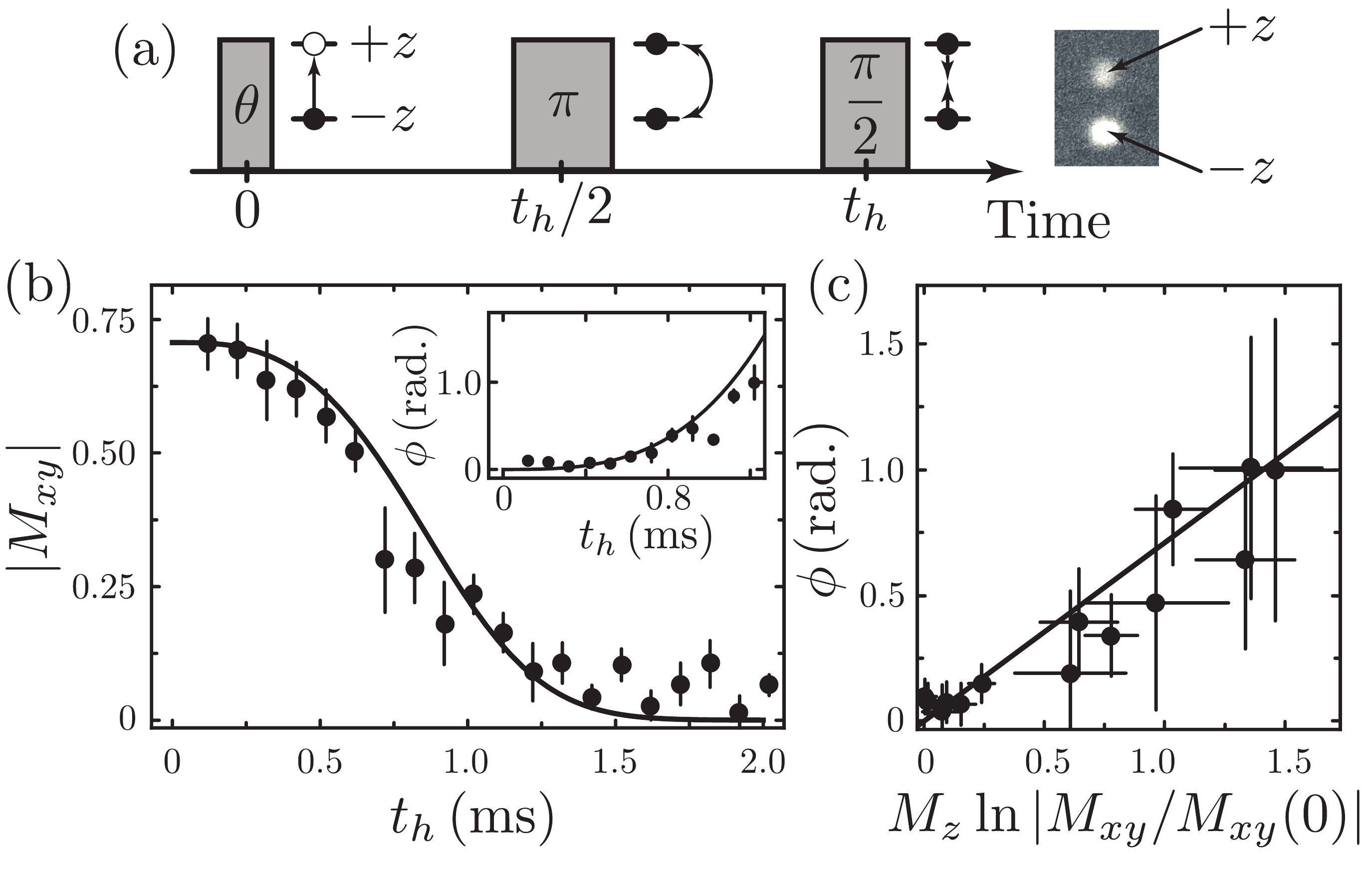}
\caption{ Magnetization dynamics. (a) The time sequence used to measure the magnetization dynamics is a simple spin-echo sequence which allows us to measure (b) the amplitude and phase (inset) of the ensemble-averaged transverse magnetization. Populations are measured with absorption imaging after Stern-Gerlach separation \cite{SM}. Data is shown for $\theta=0.25\pi$, which prepares $M_z=-0.71$. (c) $\gamma$ is found from the slope of $\phi(t_h)$ vs. $M_z \ln{| M_{xy}/M_{xy}(0) |}$. 
\label{fig:magdyn}}
\end{figure}

Transport of local magnetization $\svec{M} = \langle M_x, M_y, M_z \rangle$ occurs through a spin current $\svec{J}_j$ that can be decomposed into a longitudinal component ($\svec{J}_j^{\parallel}\parallel \svec{M}$) and a transverse component ($\svec{J}_j^{\perp}\perp \svec{M}$), where bold letters indicate vectors in Bloch space and the subscript $j \in \{1,2,3\}$ denotes spatial direction. Our measurements follow a standard spin-echo protocol \cite{Hahn:1950tv,*Carr:1954wf,*Torrey:1956tk} that initiates a purely transverse current. In the hydrodynamic regime, $\svec{J}_j^{\perp}$ is the sum of a dissipative term $- D_{\rm eff}^\perp \grad \svec{M}$ and a reactive term  $- \gamma \svec{M} \times D_{\rm eff}^\perp \grad \svec{M}$, where $D_{\rm eff}^\perp = D_0^\perp/(1+\gamma^2 M^2)$ is the effective transverse diffusivity, and $D_0^\perp$ is the bare diffusivity \cite{LR:1968,*Leggett:1970}. The parameter $\gamma$ quantifies the precession of spin current about the local magnetization, which slows demagnetization -- a phenomenon known as the Leggett-Rice effect.

Dynamics are initiated by a resonant radio-frequency (rf) pulse with area $\theta$, which creates a superposition of $\down$ and $\up$ and thus a magnetization $M_z = -\cos(\theta)$ and $M_{xy} \equiv M_x + i M_y = i\sin(\theta)$. The field gradient causes a twisting of the $xy$-magnetization into a spiral texture. The gradient in the direction of $\svec{M}$ drives a transverse spin current $\svec{J}_1^{\perp}$, which tends to relax $M_{xy} \to 0$, while $M_z$ is conserved. These dynamics are described by~\cite{LR:1968,*Leggett:1970}
\begin{equation} \label{eq:Mperp}
	\partial_t M_{xy} = -i \alpha x_1 M_{xy} + D_{\rm eff}^\perp (1+i \gamma M_z) \nabla^2_1 M_{xy} 
\end{equation}
where $\alpha = B' \Delta \mu / \hbar$, and $\Delta \mu$ is the difference in magnetic moment between $\up$ and $\down$. The solutions of Eq.~(\ref{eq:Mperp}) depend on a dimensionless time $R_M t_h$, where $t_h$ is the total hold time between the initialization pulse and final read-out pulse and $R_M \equiv (D_0^\perp \alpha^2)^{1/3}$ \cite{SM}. In our typical conditions, $R_M^{-1}$ is on  the order of $1\,$ms.

We measure the vector magnetization using a spin-echo sequence as shown in Fig.~\ref{fig:magdyn}(a). A $\pi$ pulse at time $t_h/2$ reverses all $M_{xy}$ phases, so that evolution in the presence of $B'$ causes an untwisting of the spiral magnetization texture. The final $\pi/2$ pulse is applied with a variable phase lag, so that the final populations in $\updown$ can be used to fully characterize the direction $\phi=\arg{\left( M_{xy}/i \right)}$ and the magnitude $|M_{xy}|$ of the transverse magnetization. 
 
Figure \ref{fig:magdyn}(b),(c) shows an example of $|M_{xy}(t_h)|$ and $\phi(t_h)$, for an initial pulse angle $\theta=0.25\pi$. The solution of Eq.~(\ref{eq:Mperp}) for $\gamma \neq 0$ gives $\phi = \gamma M_z\ln|M_{xy}/M_{xy}(0)|$ for all $t_h$, and thus $\gamma$ is found by linear regression on data such as Fig.~\ref{fig:magdyn}(c). Then, $R_M$ (and from it $D_0^\perp$) is determined by a nonlinearfit to $|M_{xy}(t_h)|$ data, again using an analytic solution of Eq.~(\ref{eq:Mperp}). $M_{xy}(0)$ and $B'$ are independently calibrated \cite{SM}. 

For the data shown in Fig.~\ref{fig:magdyn}, at $\lnkfa=0.13(3)$ and $\TTFi = 0.36(4)$, we find $D_0^\perp = 2.3(3) \hbar/m$ and $\gamma = 0.6(1)$. These best-fit transport coefficients are understood as an average both over the ensemble of 2D systems, and over the dynamical changes in the cloud, discussed below. At strong interaction when the mean free path $\sim1\mu$m is much smaller than the Thomas-Fermi length and the typical minimal spin-helix pitch $\sim5\mu$m, we expect that the trap averaged transport coefficients are close to the homogeneous values. In this regime the dynamics are essentially local \cite{Enss:2015er}.

\begin{figure}[b!]
\includegraphics[width=\columnwidth]{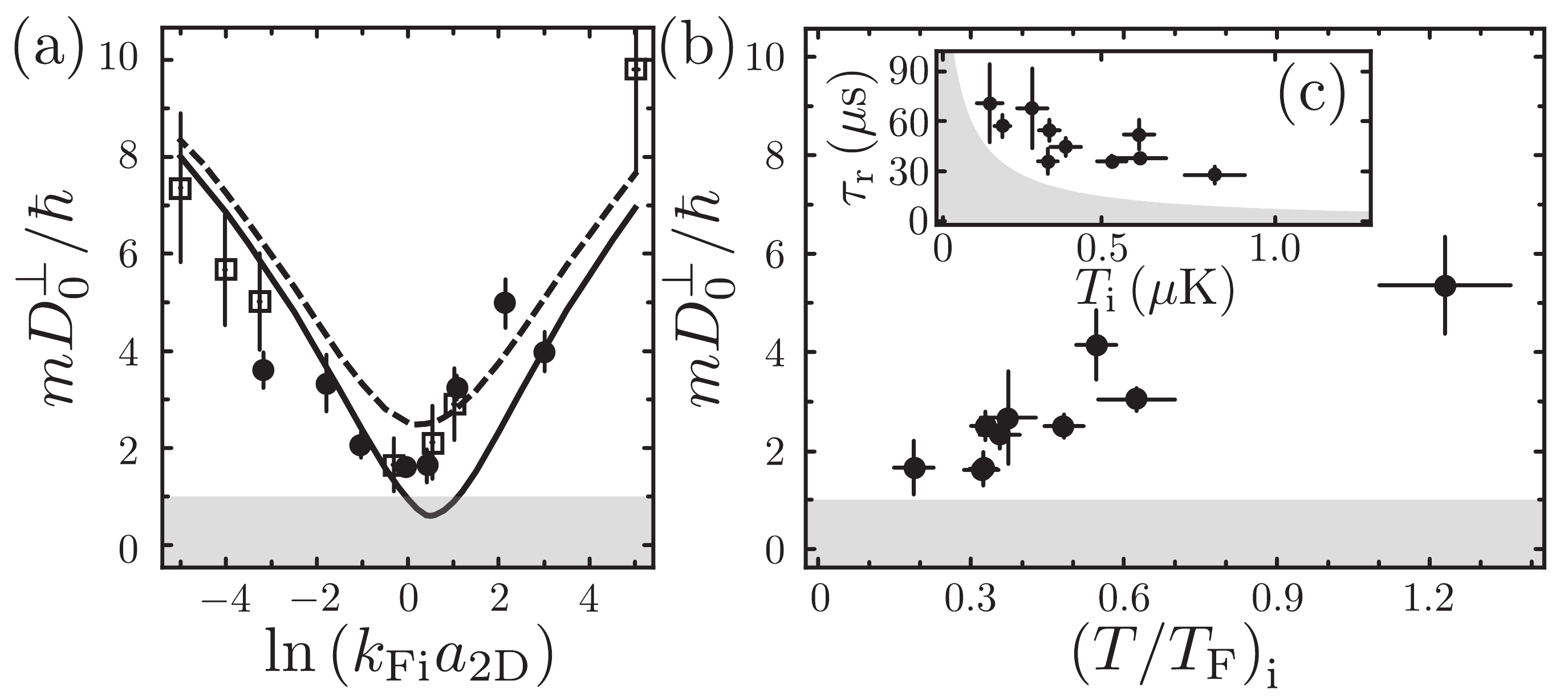}
\caption{ Transverse spin diffusivity. (a) $D_0^{\perp}$ versus interaction strength with $\TTFi=0.31(2)$ (black circles) and $\TTFi=0.21(3)$ (open squares). Each data point corresponds to a complete data set as shown in Fig.~\ref{fig:magdyn}. The lines are predictions for $T/T_\mathrm{F}=0.3$ by a kinetic theory, as described in the text. (b) $D_0^{\perp}$ versus initial reduced temperature $\TTFi$ at $\lnkfa=-0.1(2)$. (c) Local relaxation rate $\taur$ estimated as $D_0^{\perp}/v_T^2$. Shaded regions show $D_0^{\perp} < \hbar/m$ in (a,b), and $\taur < \hbar/k_B T$ in (c). Data are consistent with the conjectured quantum bound, which would exclude the shaded areas on all plots.
\label{fig:D0}}
\end{figure}

We search for conditions that minimize $D_0^{\perp}$ by repeating this characterization of $M_{xy}(t_h)$ at various interaction strengths and initial temperatures. Figure~\ref{fig:D0}(a) shows that $D_0^{\perp}$ is smallest when $-0.5 \lesssim \lnkfa \lesssim +0.5$, i.e., where $a_\mathrm{2D}$ is comparable to $k_{\rm{F}}^{-1}$. This condition can be understood by considering the 2D scattering amplitude in vacuum: $f(k)=2 \pi/[ -\ln(k a_\mathrm{2D}) + i \pi/2 ]$ \cite{Engelbrecht:1992eo,Petrov:2001gx,Bloch:2008gl,Levinsen:2015dt}, which gives a maximal (unitary) cross-section $4/k$ at $k a_\mathrm{2D}=1$. Even though our Fermi gas has a distribution of relative momenta $k$, the average cross-section at low temperature can be estimated by replacement of $k$ with $k_{\rm{F}}$, due to the logarithmic dependence of $f$ on the energy of collision. In other words, corrections to the unitary scattering cross section are only logarithmic \cite{Hofmann:2011ds,Langmack:2012hr,Schafer:2012,Bruun:2012,*Enss:2012a,Hofmann:2012ic,Taylor:2012if}, which explains the qualitative similarity of Fig.~\ref{fig:D0}(a) to prior 3D measurements \cite{Trotzky:2015fe}.

The lines on Fig.~\ref{fig:D0}(a) show a kinetic theory both with and without medium scattering (solid and dashed lines, respectively) calculated in the $|\svec{M}|\to1$ limit \cite{Enss:2013ti,Enss:2015er}. The model also accounts for inhomogeneities in the following way: first, the collision integral is solved to compute the transverse spin diffusion time and LR parameter for a 2D homogeneous system with the same spin density and temperature as the trap center \cite{Jeon:1989,Enss:2013ti}. Next, these parameters are used to solve the Boltzmann equation for the position-dependent spin density in the full trapping potential for each 2D gas in the ensemble \cite{Enss:2015er}. Finally, the average magnetization dynamics is analyzed using Eq.~(\ref{eq:Mperp}). This procedure predicts a minimal $D_0^{\perp}$ slightly shifted from the observed minimum; but its results agree well with the increase of $D_0^{\perp}$ in the weakly interacting regime. This gives us confidence that inhomogeneity effects are well understood.

The lowest observed diffusivity is $D_0^\perp = 1.7(6)\hbar/m$, at $\TTFi=0.19(3)$ and $\lnkfa=-0.1(2)$. The effect of temperature is shown in Fig.~\ref{fig:D0}(b) and by data sets in Fig.~\ref{fig:D0}(a) taken at two temperatures. In all cases, our data supports the conjectured bound $D_0^{\perp} \gtrsim \hbar/m$. 

Assuming that magnetization perturbations propagate at $v_T\sim\sqrt{k_B T/m}$, one can estimate the local relaxation time $\taur$ with $D_0^\perp/v_T^2$. Figure~\ref{fig:D0}(c) compares this time to the bound $\hbar/k_B T$. Another estimate of the relaxation time would use the Fermi velocity $v_{\rm{F}}$, as $\taur \sim 2 D_0/v_\mathrm{F}^2$, which is the correct scaling for mean free time in imbalanced Fermi liquids at low temperature \cite{Jeon:1989,LR:1968,*Leggett:1970,SM}. This yields $\taur \sim 20\,\mu$s at the minimum observed diffusivity, again on the order of $\hbar/k_B T$. In sum, a 2D Fermi gas with $a_\mathrm{2D} k_\mathrm{F} \sim 1$ seems to saturate, but not violate, the Planckian bound $\taur^{-1} \lesssim k_B T/\hbar$ at the lowest temperatures probed here.

\begin{figure}[tb!]
\includegraphics[width=\columnwidth]{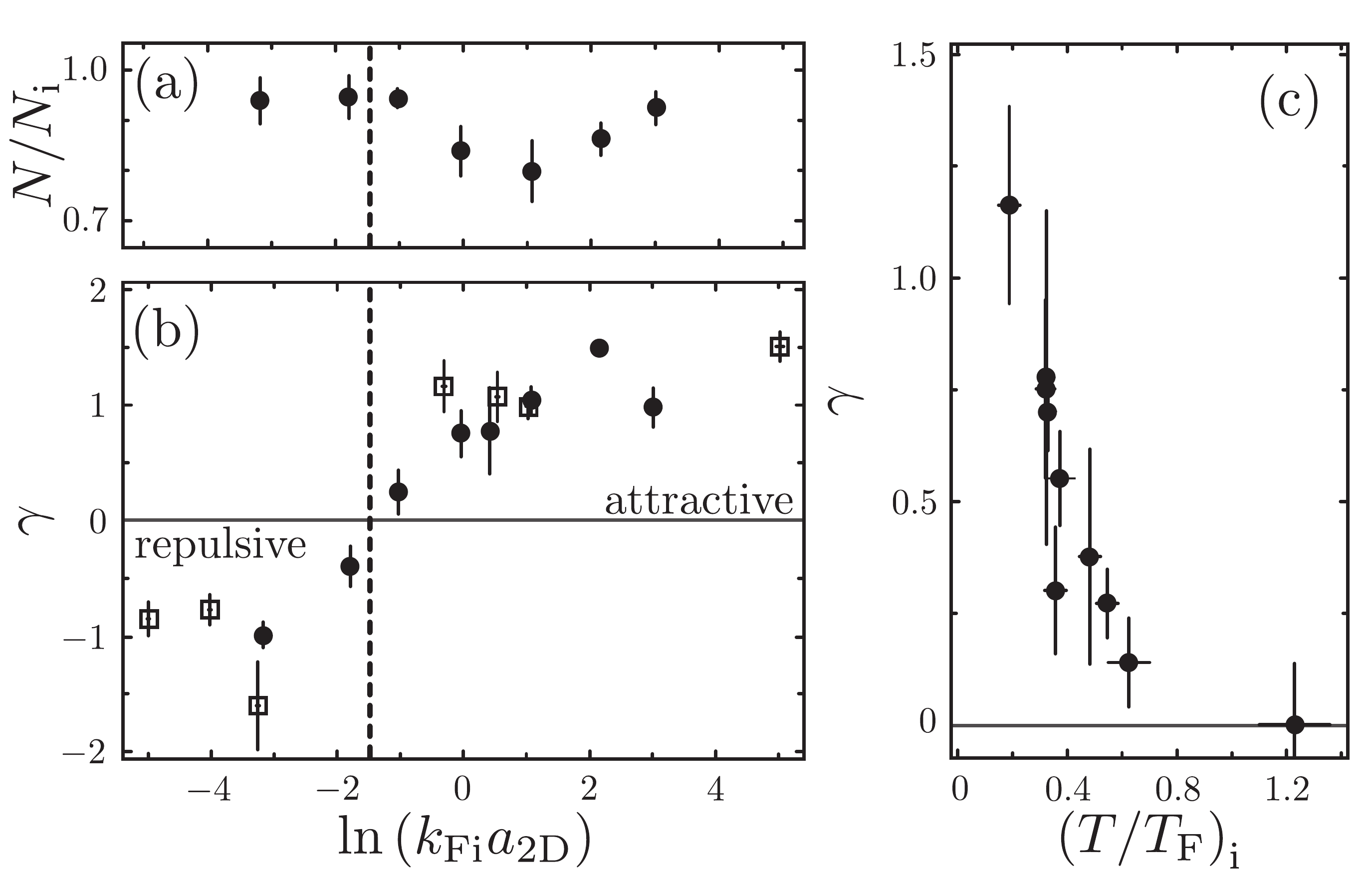}
\caption{Change in the sign of interaction. (a) Fraction of atoms remaining at $t_h=3.5$\,ms. (b) $\gamma$ versus interaction strength, with markers as in Fig.~\ref{fig:D0}(a). (c) $\gamma$ versus initial reduced temperature $\TTFi$, at $\lnkfa=-0.1(2)$. The change in sign of $\gamma$, at $\lnkfa \approx -1$, is associated with the onset of a pairing instability.
\label{fig:LRE}}
\end{figure}

Figures \ref{fig:LRE}(b) and \ref{fig:LRE}(c) summarize measurements of $\gamma$ across a wide range of interaction strengths and temperatures. There are two implications of these data. First, system-wide demagnetization is slowed by spin current precession, to an apparent diffusivity $D_{\rm eff}^\perp$, which is initially $D_0^\perp/(1+\gamma^2)$ for a fully polarized cloud. This is a reasonable quantity to compare to the ``$D_s^\perp$'' measured in Ref.~\cite{Kohl2013} to be $6.3(8) \times 10^{-3} \hbar/m$ at minimum. In similar conditions, we instead find $D_{\rm eff}^{\perp} = 7(3) \times 10^{-1} \hbar/m$. In both works, diffusivity is observed to be minimal near $\lnkfa = 0$, and to double between $\lnkfa \approx 0$ and $\lnkfa \approx 1$. However, we cannot explain the hundred-fold difference in scale.

The second implication of $\gamma$ is to reveal the sign of the interaction between the spin current and the local magnetization \cite{LR:1968,*Leggett:1970,Laloe:1982wc,Miyake:1985hz}. When $\gamma < 0$, as we observe for $\lnkfa \lesssim -1.5$ [see Fig.~\ref{fig:LRE}(b)], interactions are repulsive, whereas when $\gamma > 0$, as we observe for $\lnkfa \gtrsim -1.5$, interactions are attractive. 
Associated with the sign change of $\gamma$ is the onset of a pairing instability, since both are related to the sign change of the real part of the low-energy scattering $\T$ matrix \cite{Enss:2013ti,Trotzky:2015fe,Pekker:2011fh,*Sodemann:2012eq}. We find indirect evidence for this from atom loss [see Fig.~\ref{fig:LRE}(a)], since Feshbach dimers are a precursor to formation of deeply bound molecules \cite{Zhang:2011ga}, which are lost from the trap. In 3D, this loss rate is higher on the repulsive side of unitarity; but in 2D, we observe the strongest loss on the attractive side, at $\lnkfa \sim 1$ \cite{Pietila:2012uk}. We discuss this further below.

\begin{figure}[tb!]
\includegraphics[width=\columnwidth]{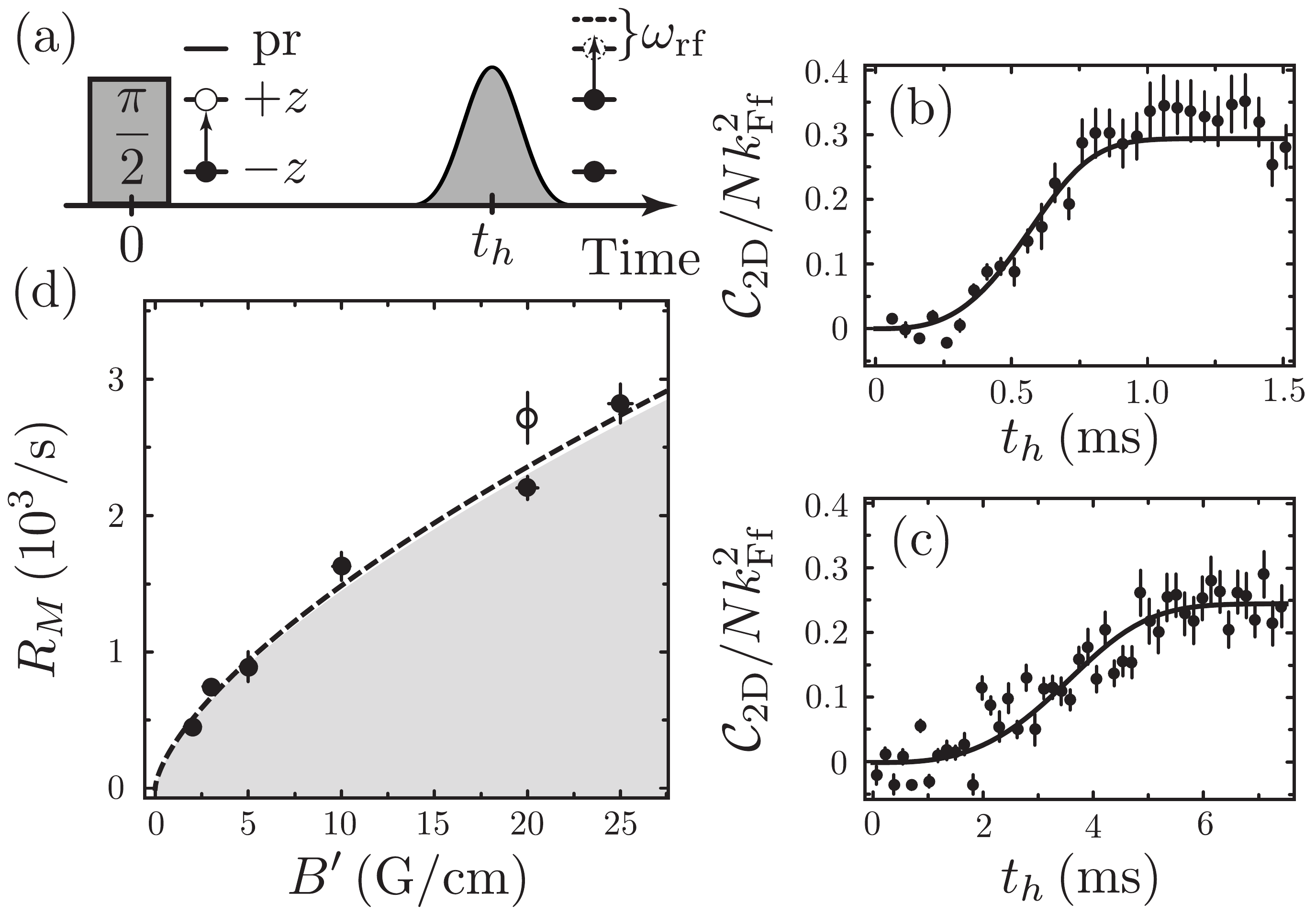}
\caption{Contact Dynamics at $\lnkfa=0.35(5)$ and $(T/T_\mathrm{F})_\mathrm{i} = 0.31(2)$. (a) $\Contact$ is measured after a hold time $t_h$ by a pulse detuned by $\omega_{\rm{rf}}$ from the $\up$-to-$\probe$ transition. (b) Contact growth for $B'=25$\,G/cm. (c) Contact growth for $B'=2$\,G/cm. (d) The best-fit $R_M$ determined from contact growth (black points), versus $B'$. The shaded region corresponds to $R_M$ with $D_0^\perp < \hbar/m$. The open point indicates $R_M$  from $\svec{M}$ dynamics at 20\,G/cm. The dashed line shows the best-fit diffusivity $D_0^{\perp}=1.1(1)\hbar/m$.
\label{fig:contact}}
\end{figure}

One consequence of demagnetization is a cloud-wide redistribution of energy. For a 2D harmonically trapped Fermi gas, the virial relation is \cite{Werner:2008is,*Valiente:2011}
\begin{equation} \label{eq:virial}
V=\frac{1}{2}E+\frac{\hbar^2}{8\pi m}\Contact
\end{equation} 
where $V$ is the total potential energy, $E$ is the total energy, and $\Contact$ is the (extensive) 2D contact \cite{Tan:2008ch,*Braaten:2008tc,*Zhang:2009kq,*Combescot:2009gw,*Werner:2012um,Werner:2008is,*Valiente:2011}.
Even though the trap explicitly breaks scale invariance, an SO$(2,1)$ dynamical symmetry survives at the mean-field level \cite{Pitaevskii:1997gk}, but is broken by a quantum anomaly whose expectation value is $\Contact$ \cite{Hofmann:2012ic}. $E$ is conserved in this isolated system, however $\Contact$ increases from zero for the non-interacting initial state to a finite positive value for the final state. This implies that $V$ must also increase, which in turn dictates an increase in the rms cloud size: $V/N=\frac{1}{2}m(\omega_1^2\langle x_1^2\rangle+\omega_2^2\langle x_2^2\rangle)$. 

Using rf spectroscopy, we measure $\Contact$ throughout the demagnetization dynamics. The protocol is as described in Ref.~\cite{Bardon:2014} and depicted in Fig.~\ref{fig:contact}(a). The dynamics are initiated with a $\theta=\pi/2$ pulse and the sample is probed with a spectroscopic pulse that couples the states $\up$ and $\probe$ after a hold time $t_h$. The transfer rate of population to state $\probe$ is measured as a function of the detuning $\omega_{\rm{rf}}$ from the bare spin-flip resonance, and is known to scale with $\Contact\omega_{\rm{rf}}^{-2}$ in the limit $\omega_{\rm{rf}}\gg E_{\rm{F}}$ \cite{Braaten2010,Langmack:2012hr,Randeria2010,Frohlich:2012ic}. We compensate for final-state interactions between the $\probe$ atoms and $\updown$ atoms in our analysis \cite{SM, Langmack:2012hr}.

At $\ln{(k_{\rm{Ff}}a_{\rm{2D}})}=0.00(5)$ we find that the contact rises from zero to $\Contact/N =0.28(3) k_{\rm{Ff}}^2$, where $\kff^2=\kfi^2/2$ after complete depolarization. Using Eq.~(\ref{eq:virial}), one finds $V-E/2 = 0.022(2) E_\mathrm{Ff}$ per particle. In contrast, for a 3D gas at unitarity, there is no correction to the virial: $V-E/2$ is proportional to $\mathcal{C}_\mathrm{3D}/a_\mathrm{3D}$, and goes to zero when $a_\mathrm{3D}^{-1} \to 0$.

A final thermodynamic transformation accompanying demagnetization is a temperature rise due to the combination of increased spin entropy and decreased occupation of the Fermi sea \cite{Ragan:2002tx}. For an initial temperature of $0.3(1)\,T_\mathrm{Fi}$ and a $\pi/2$ pulse,  we observe $T_{\rm{f}}=0.7(2)\,T_\mathrm{Ff}$ near $\ln{(k_{\rm{Ff}}a_{\rm{2D}})}=0$. Due to the released attractive interaction energy, this temperature rise is larger than the $\Delta(T/T_\mathrm{F})\approx 0.25$ one would expect from demagnetization of an ideal gas. However the observed heating is three times smaller than the $\Delta(T/T_\mathrm{F})\approx 2.2$ that is predicted by matching initial energy and number to the equilibrium 2D equation of state \cite{Bertaina:2011kk,*Bauer:2014fa,*Shi:2015ku,*Anderson:2015wj,*Galea:2016ce,Fenech:2016cv,Boettcher:2016fz}.

One interpretation of these observations is that few or no dimers are formed during demagnetization. 
This is certainly true on the repulsive side ($a_{\rm{2D}} < k_{\rm{F}}^{-1}$) of unitarity, where the system is not a dimerized superfluid as it would be in the ground state. But even at unitarity, measurements of $T$ and $\Contact$ suggest that the system remains in the {\em upper energetic branch}. The value of $\Contact/N k_\mathrm{F}^2$ we observe is roughly twenty times smaller than the contact strength in an equilibrium mixture at $\ln{(k_{\rm{F}}a_{\rm{2D}})} = 0$ \cite{Frohlich:2012ic,Bertaina:2011kk,*Bauer:2014fa,*Shi:2015ku,*Anderson:2015wj,*Galea:2016ce}. The equilibrium contact is primarily due to a mean-field dimer contribution $\mathcal{C}_0 \approx 4 N k_F^2$. Without dimers, the contact in the upper branch would be due to short-range correlations of unbound atoms, and in fact the value we observe is comparable to $\Contact - \mathcal{C}_0$ in the lower branch
 \footnote{The peak ``many-body contact'' density in the lower branch is roughly $0.055 k_F^4$ \cite{Bertaina:2011kk,*Bauer:2014fa,*Shi:2015ku,*Anderson:2015wj,*Galea:2016ce}, and when divided by density $n = k_F^2/2\pi$ gives a bulk uniform contact $\Contact - \mathcal{C}_0 \approx 0.35 N k_F^2$.}.
Unlike in 3D, at the unitarity point in 2D the dimer binding energy is greater than $E_\mathrm{F}$, so that an {\em attractive} upper branch is energetically well defined.

Figures \ref{fig:contact}(b),(c) show the typical dynamics we observe when measuring $\Contact(t_h)$. Due to Pauli exclusion, we can use such data to infer magnetization dynamics: pairs of fermions must have a singlet wave function to interact through an $s$-wave contact interaction. The singlet fraction can be no larger than $1-|\svec{M}|$, and would be $(1-|\svec{M}|^2)/4$ for uncorrelated spins \cite{Bardon:2014,Thekkadath:2016fs,He:2016vm}. For the $\pi/2$ initialization pulse performed here, $|\svec{M}|=|M_{xy}|$ since $M_z=0$. A direct comparison between $\svec{M}$ and $\Contact$ at $B'=20$ G/cm (see \cite{SM}) shows a correlation that lies between these two limits: $\Contact/N$ is proportional to $1-|M_{xy}(t_h)|^{1.4(2)}$. This form with $\gamma=0.71$ is used to fit $\Contact$ data for a variety of gradients [see  Fig.~\ref{fig:contact}(b),(c)] and extract $R_M$. 

Across the experimentally accessible gradients $B'$, Fig.~\ref{fig:contact}(d) shows a range of $R_M$ from $4.4(2) \times 10^2$\,s$^{-1}$ to $2.9(2) \times 10^3$\,s$^{-1}$. Throughout, $R_M$ scales with $\alpha^{2/3}$ (see dashed line) and can be explained by a single diffusivity $D_0^\perp=1.1(1)\hbar/m$. This verifies that the microscopic $D_0^\perp$ is independent of $B'$ across the accessible range, and thus independent of the pitch of the spin helix. The comparable magnitude of $D_0^\perp$ determined by two measurement techniques is also a reassuring check on the fidelity of the spin-echo sequence used in $\svec{M}$ measurements, since the measurement of $\Contact$ does not rely upon successful rephasing of the spins at the echo time.


In sum, we observe quantum-limited spin transport in 2D Fermi gases when $a_{\rm{2D}}$ is tuned to be comparable to $k_{\rm{F}}^{-1}$. We find that the conjectured lower bound $D_0^\perp \gtrsim \hbar/m$ is respected for all interaction strengths, temperatures, and applied field gradients accessible to our apparatus. This supports the generality of the bound $\taur^{-1} \lesssim k_B T/\hbar$ beyond quantum critical systems, since the finite $\Contact$ observed in this system signifies a broken scaling symmetry near unitarity.

\begin{acknowledgments}
\acknowledgments{We thank A.\ Georges, A.\ Gezerlis, M.\ K\"ohl, S.\ Sachdev, E.\ Taylor, and Shizhong Zhang for stimulating discussions. This work is supported by NSERC, by AFOSR under FA9550-13-1-0063, by ARO under W911NF-14-1-0282, and is part of and supported by the DFG Collaborative Research Centre ``SFB 1225 (ISOQUANT)''.}
\end{acknowledgments}

\bibliography{bibLR2D}

\end{document}


\title{Supplementary Material for\\
Observation of quantum-limited spin transport in strongly interacting two-dimensional Fermi gases}
\author{C.\ Luciuk}
\author{S.\ Smale}
\affiliation{Department of Physics, University of Toronto, Ontario M5S\,1A7 Canada}
\author{F.\ B\"ottcher}
\affiliation{5.\,Physikalisches Institut and Centre for Integrated Quantum Science and Technology, Universit\"at Stuttgart, D-70569 Stuttgart, Germany}
\author{H.\ Sharum}
\author{B.\ A.\ Olsen}
\author{S.\ Trotzky}
\affiliation{Department of Physics, University of Toronto, Ontario M5S\,1A7 Canada}
\author{T.\ Enss}
\affiliation{Institut f\"ur Theoretische Physik, Universit\"at Heidelberg, D-69120 Heidelberg, Germany}
\author{J.\ H.\ Thywissen}
\affiliation{Department of Physics, University of Toronto, Ontario M5S\,1A7 Canada}
\affiliation{Canadian Institute for Advanced Research, Toronto, Ontario M5G\,1Z8 Canada}

\maketitle

\renewcommand\thefigure{S\arabic{figure}}
\renewcommand\theequation{S\arabic{equation}}

\section{Experimental Methods}
\paragraph*{Sample preparation.}
We sympathetically cool fermionic spin-polarized $^{40}$K using bosonic $^{87}$Rb. Initially both species are trapped in a microfabricated magnetic chip trap where $^{87}$Rb undergoes forced rf evaporative cooling. Subsequently both species are loaded into a crossed-beam optical dipole trap with aspect ratio 6:1:1, followed by a final stage of evaporative cooling. A resonant light pulse removes any remaining $^{87}$Rb leaving a spin-polarized degenerate Fermi gas, typically with $N=4\times10^4$ atoms at a temperature $T=150$ nK in the lowest hyperfine state of the electronic ground state. The $\down$, $\up$, and $\probe$ states in the main text refer to the high-field states adiabatically connected to the low-field $m_f=-9/2,\,-7/2,\,$and $-5/2$ states of the $f=9/2$ hyperfine manifold, where $f$ and $m_f$ refer to the total angular momentum and corresponding magnetic quantum number, respectively.

Our 3D sample is adiabatically loaded into a one-dimensional optical lattice formed from a standing wave of 760.2 nm light with corresponding recoil energy $E_R/\hbar=2\pi\times8.64$ kHz. The lattice depth $V_0$ is ramped up in two stages: first to approximately 5 $E_R$ in 100 ms where the tunneling between adjacent sites is effectively frozen out, and then in 5 ms to 50 $E_R$ where we perform our experiments. The final lattice depth is determined using amplitude modulation spectroscopy to resonantly create excitations from the ground band to the second excited band of the lattice which we observe as population in the second Brillouin zone after band-mapping.

This loading procedure typically populates two hundred planes with radial trapping frequency $\omega_{1,2}=2\pi\times600(35)$ Hz and tight transverse confinement $\omega_3=2\pi\times122(35)$ kHz. In the central 2D gas we typically have 500 atoms with a global initial Fermi energy $E_{\rm{Fi}}=\hbar\omega_{1,2}(2N)^{1/2}\approx h\times19$ kHz, well within the 2D regime ($E_{\rm{F}}\ll\hbar\omega_3$).

\paragraph*{Temperature control and determination.}
We vary the temperature of our ensemble of 2D gases by varying the initial temperature before loading the optical lattice, by choosing the final trap depth of evaporation in the crossed-beam optical dipole trap. Adiabatic loading of the spin-polarized gas into the optical lattice causes an increase in the reduced temperature $T/T_{\rm{F}}$ due to the modified density of states (DoS). While the (semiclassical) DoS for a three-dimensional harmonic oscillator is $\propto E^2$, the introduction of the optical lattice along one spatial direction changes the power law to $\propto E^{3/2}$. 

\begin{figure}[t!]
\includegraphics[width=\columnwidth]{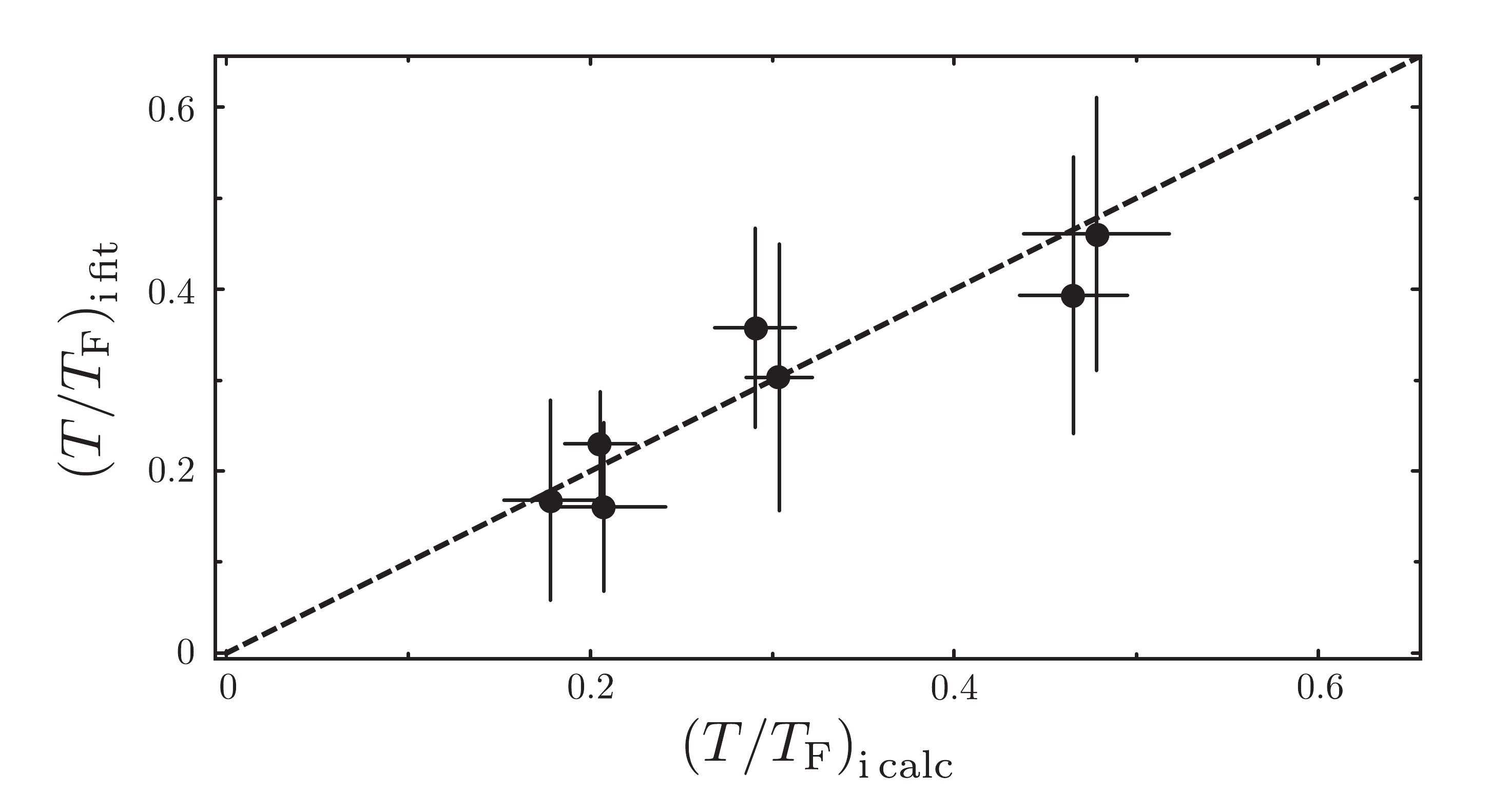}
\caption{Thermometry. A comparison of the fitted temperature $(T/T_{\rm{F}})_{\rm{i\,fit}}$ and calculated ensemble average temperature $(T/T_{\rm{F}})_{\rm{i\,calc}}$. The vertical error bars show statistical uncertainty from 10-20 images. The dashed line has a slope of unity.}
\label{fig:temp}
\end{figure}

Given the temperature after evaporation, atom number, and measured trap parameters, we calculate the temperature expected in the lattice. Motivated by the loading procedure, we assume isentropic loading into a lattice depth of $5\,E_R$. The final ramp to $50\,E_R$ is assumed to cause only a change in the confinement of the isolated 2D gases, without changing the number or reduced temperature in each. Figure~\ref{fig:temp} compares this calculated ensemble-averaged reduced temperature to measurements of the same using absorption images taken after release from the deep lattice trap and subsequent expansion during time of flight (see Fig.~\ref{fig:temp}). They agree within their statistical uncertainties, providing evidence that the loading process is indeed adiabatic. 

The $\TTFi$ given in the main text uses the calculated temperature and number in the central 2D system, from which $E_{\rm{Fi}} = \hbar \omega_{1,2} (2 N)^{1/2}$ and $k_{\rm{Fi}}^2 = 2 m E_{\rm{Fi}}/\hbar^2$. The uncertainties for $\TTFi$ in the main text are determined by repeating the calculation for the observed distribution of atom number and temperature, which appear both as shot-to-shot fluctuations and as drifts during each data run (approximately 3 hours).

\paragraph*{Field and gradient control.}
We control the magnetic field and its gradients using a combination of magnetic field coils and wires on the atom chip $\approx 200 \rm{\mu m}$ from the atoms. We tune the field $|{\bf{B}}|$ near 202.1 G where the states $\updown$ undergo a magnetic Feshbach resonance. We stabilize the field to about 1 part in $10^5$. We calibrate the field by measuring the $\down$ to $\up$ transition frequency and converting the frequency to magnetic field using the Breit-Rabi relation. During a measurement the field drifts by as much as 0.02 G which, when combined with the number and temperature uncertainty, gives an uncertainty in $\ln{(k_{\rm{F}}a_{\rm{2D}})}$ of $0.04$.

We vary the gradient $B'=|\langle\nabla_1B,\nabla_2B,\nabla_3B\rangle|$ by changing the sum and difference of currents through parallel wires on the atom chip. Unless otherwise stated in the main text, we choose a gradient of $\nabla_1B=20.3(2)$ G/cm and $\nabla_2B=0.3(6)$ G/cm measured by repeating spectroscopy measurements on a cloud translated by piezo-actuated mirrors on the trapping beams. There is a residual gradient $\nabla_3B\approx-2.7$ G/cm that we cannot control, however, this is in the same direction as the tight confinement and does not contribute to the demagnetization dynamics. 

\paragraph*{Dimer binding energy.}
For low-energy scattering in a quasi-2D system the 2D scattering length is given by
\begin{equation}
a_\mathrm{2D} = \ell \sqrt{\pi/B} \exp{[-\sqrt{\pi/2} \ell/a]}
\end{equation}
where $B\approx0.905$, $\ell$ is the harmonic oscillator length in the direction of strong confinement, and $a$ is the 3D scattering length. This result is valid across the three-dimensional resonance. The dimer binding energy is determined from solutions of a transcendental equation (c.f. Ref.~\cite{Levinsen:2015dt}). When $\ell/a\lesssim -1$ the dimer binding energy matches the 2D expression $\hbar^2/m a_\mathrm{2D}^2$. For $\ell/a \gtrsim 1$, dimers are three-dimensional and have a binding energy $\approx \hbar \omega_3/2 + \hbar^2/m a^2$ \cite{Petrov:2001gx,Bloch:2008gl,Levinsen:2015dt}.

\paragraph*{Imaging.}
Our imaging scheme allows us to count the population of atoms in two $m_f$ states. For the demagnetization dynamics we image atoms in states $\down$ and $\up$, while for the spectroscopy measurements we image atoms in states $\up$ and $\probe$. We achieve this population counting by applying a Stern-Gerlach pulse to separate the trapped spin states, rf state manipulation during time-of-flight in a residual gradient, and finally absorption imaging of the $|f=9/2,\,m_f=-9/2\rangle$ to $|f'=11/2,\,m_f'=-11/2\rangle$ cycling transition.

\paragraph*{Fitting of magnetization dynamics.}
We measure the magnetization decay for various initial amplitudes of the transverse magnetization $|M_{xy}(0)|=\sin{\theta}$ using a $\theta-\pi-\pi/2$ pulse sequence. To minimize the duration of data sets we fix $\theta\approx0.25\pi$ calibrated by optimizing a $\pi$ pulse and then decreasing the duration of the pulse by $1/4$. The first two pulses have the same phase while the last pulse has a variable relative phase lag. An oscillation in the population is generated by varying the phase lag from which the magnitude $|M_{xy}|$ and phase $\phi=\arg{(M_{xy}/i)}$ of the transverse magnetization can be measured. We determine the amplitude and phase of the transverse magnetization from a sinusoidal fit to this oscillation.

We are sensitive to the relative frequency between the drive and the atomic frequency with a precision of $\sim 1/t_h$, where $t_h$ is the hold time. For $t_h\ge1.5$ ms we find that our field stability is insufficient to produce a reproducible phase, and thus to determine $\gamma$ or the bare $D_0^\perp$. To avoid this, we work at a sufficiently high magnetic field gradient that the demagnetization time is less than 1.5\,ms. As described in the main text, lower field gradients are explored using $\Contact$ dynamics, which is insensitive to uniform field fluctuations.

For each data set \{$|M_{xy}(t_h)|$, $\phi(t_h)$\} we determine $D_0^{\perp}$ and $\gamma$ using the solution to Eq.~(1) given by
\begin{align}
|M_{xy}(t_h)| &= |M_{xy}(0)|\sqrt{ \frac{1}{\eta}{\mathcal{W}}\left( \eta\exp{ \left[ \eta-\frac{(R_Mt_h)^3}{6\left( 1+\gamma^2M_z^2 \right)} \right] } \right) } \nonumber \\
\phi(t_h) &= \gamma M_z \ln{\left( \frac{|M_{xy}(t_h)|}{|M_{xy}(0)|} \right)} \label{eq:amp}
\end{align}
where $\alpha=B' \Delta\mu/\hbar$, $\eta=\gamma^2|M_{xy}(0)|^2/(1+\gamma^2M_z^2)$, and ${\mathcal{W}}(z)$ is the Lambert-W function. The characteristic time scale for the dynamics is $R_M^{-1} \equiv (D_0^\perp \alpha^2)^{-1/3}$. We plot $\phi(t_h)$ versus $M_z\ln{(|M_{xy}(t_h)|/|M_{xy}(0)|)}$ as in Fig.~1(c) of the main text. A linear fit to this data determines $\gamma$. We then fix $\gamma$ and fit $|M_{xy}(t_h)|$ to determine $D_0^{\perp}$. After determination of $\gamma$, the only effect of $D_0^{\perp}$ is to rescale time; in other words, for any $D_0^{\perp}$, $\alpha$, or $t_h$, Eq.~(\ref{eq:amp}) is a universal function of $R_M t_h$ and $\gamma$.

\section{Fermi liquid theory in 2D}
For a dilute, weakly repulsive Fermi gas in two dimensions, the ground state is a stable Fermi liquid \cite{Engelbrecht:1992eo} and the Landau interaction $f$ function is given to second order in the coupling $g=-1/\ln(k_{\rm{F}}a_\text{2D})$ by
\begin{align}
  \label{eq:Atheta}
  F^\text{s}(\theta)
  & = g+g^2\left(2+\ln\left\lvert
    \cos\frac{\theta}{2}\right\rvert\right) \\
  F^\text{a}(\theta)
  & = -g-g^2\ln\left\lvert\cos\frac{\theta}{2}\right\rvert.
\end{align}
The first angular Fourier coefficients defined by $F^\text{s,a}(\theta)=\sum_m F_m^\text{s,a} e^{im\theta}$ are
\begin{align}
  \label{eq:Fs}
  F_0^\text{s} & = g+g^2(2-\ln2), & F_1^\text{s} & = g^2/2, \\
  \label{eq:Fa}
  F_0^\text{a} & = -g+g^2\ln2, & F_1^\text{a} & = -g^2/2.
\end{align}
The effective mass $m^*/m=1+F_1^\text{s}=1+g^2/2$ and the spin susceptibility $\chi/\chi_0 = (1+F_1^\text{s})/(1+F_0^\text{a}) = 1+g+g^2(3/2-\ln2)$.  The transverse spin diffusivity in 2D at low temperature is given by
\begin{align}
  \label{eq:4}
  D_0^\perp = \frac{\chi_0}{\chi}\, \frac{\tau_\perp \varepsilon_{\rm{F}}}{m^*}
  = (1+F_0^\text{a})\frac{\tau_D v_{\rm{F}}^2}{2},
\end{align}
where $\varepsilon_{\rm{F}}$ is the Fermi energy and $v_{\rm{F}}$ is the Fermi velocity, or conversely the transverse scattering time is
\begin{align}
  \label{eq:tauperp}
  \tau_D = \frac{2D_0^\perp}{v_{\rm{F}}^2(1+F_0^\text{a})}.
\end{align}
For $(1+F_0^\text{a})$ of order unity, this relation agrees with the $v_{\rm{F}}$-based estimate of $\tau_\mathrm{r} \sim \tau_D$ given in the main text. 
The spin-rotation parameter can be written in terms of the scattering time and a mean-field interaction $W$ between unlike spins \cite{Enss:2015er},
\begin{align}
  \gamma = -\frac{\tau_D Wn}{\hbar}.
\end{align}
Following the notation in \cite{Trotzky:2015fe}, the relation of $\gamma$ to the parameter $\mu$ used in the liquid $^3$He literature is $\gamma = n \mu/2$ where $n$ is the total number density.

\begin{figure}[b!]
\includegraphics[width=\columnwidth]{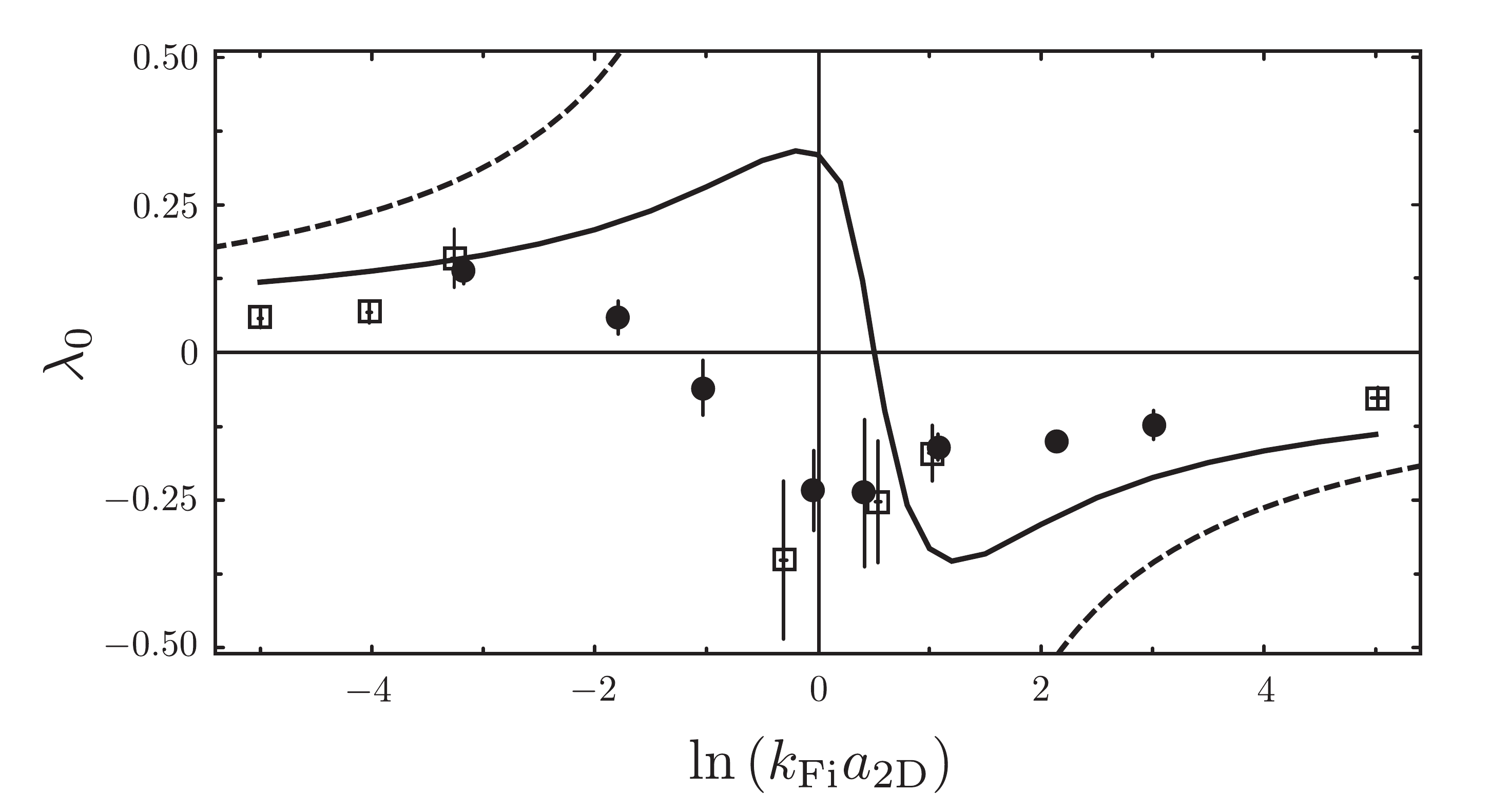}
\caption{Comparison to FLT. The effective interaction $\lambda_0$ is plotted for a range of temperatures and interaction strengths. Data points show $\lambda_0=-{\hbar\gamma}/{(2m D_0^\perp)}$ for $\TTFi=0.31(2)$ (black circles) and $\TTFi=0.21(3)$ (open squares). Dashed lines show $\lambda$ predicted by weak coupling theory, Eq.~(\ref{eq:3}); solid lines show a $\T$-matrix calculation for the trapped gas. 
\label{fig:lambda0}}
\end{figure}

The ratio $\gamma/D_0^\perp$ eliminates the scattering time and allows one to define the dimensionless effective interaction $\lambda$ in terms of the measured quantities $\gamma$ and $D_0^\perp$,
\begin{align}
  \lambda = -\frac{\hbar\gamma}{2m^*D_0^\perp}.
\end{align}
From the definition of the mean-field interaction $W$ \cite{Enss:2015er} one can express $\lambda$ in terms of the Landau parameters,
\begin{align}
  \label{eq:lambda}
  \lambda 
  = \frac{g(0)W}{1+F_0^\text{a}}
  = \frac{1}{1+F_0^\text{a}} - \frac{1}{1+F_1^\text{a}}
\end{align}
where $g(0)$ denotes the density of states per spin state.  To second order in the perturbative weak-coupling expansion \eqref{eq:Fa} we find
\begin{align}
  \lambda & = g - g^2(\ln2-1/2) + \mathcal O(g^3) \\
  & = -\frac{1}{\ln(k_{\rm{F}}a_\text{2D})-(\ln2-1/2)}
    + \mathcal O(\lvert\ln(k_{\rm{F}}a_\text{2D})\rvert^{-3}).\label{eq:3}
\end{align}
The divergence near $\ln(k_{\rm{F}}a_\text{2D}) = \ln2-1/2 \approx 0.2$ signals the breakdown of the perturbative expansion. The true solution should exhibit a zero crossing as the interacting gas crosses over from effective repulsive interactions $\lambda>0$ at $\ln(k_{\rm{F}}a_{\rm{2D}})\lesssim -1$ to effective attractive interactions $\lambda<0$ at $\ln(k_{\rm{F}}a_{\rm{2D}}) \gtrsim 1$. 

Figure~\ref{fig:lambda0} compares the FLT prediction of $\lambda$, the $\T$-matrix calculation of $\lambda$, and the measured values of $\lambda_0=(m^*/m)\lambda$ (which needs no knowledge of mass renormalization). Near $\lnkfa\approx0.2$, both experiment and the $\T$-matrix calculation give a finite value in contrast to the divergence predicted by the weak-coupling expansion. We note that the largest measured magnitude of $\lambda_0$ in 2D is nearly twice as large as the same dimensionless quantity in 3D \cite{Trotzky:2015fe}.

\begin{figure}[b!]
\includegraphics[width=\columnwidth]{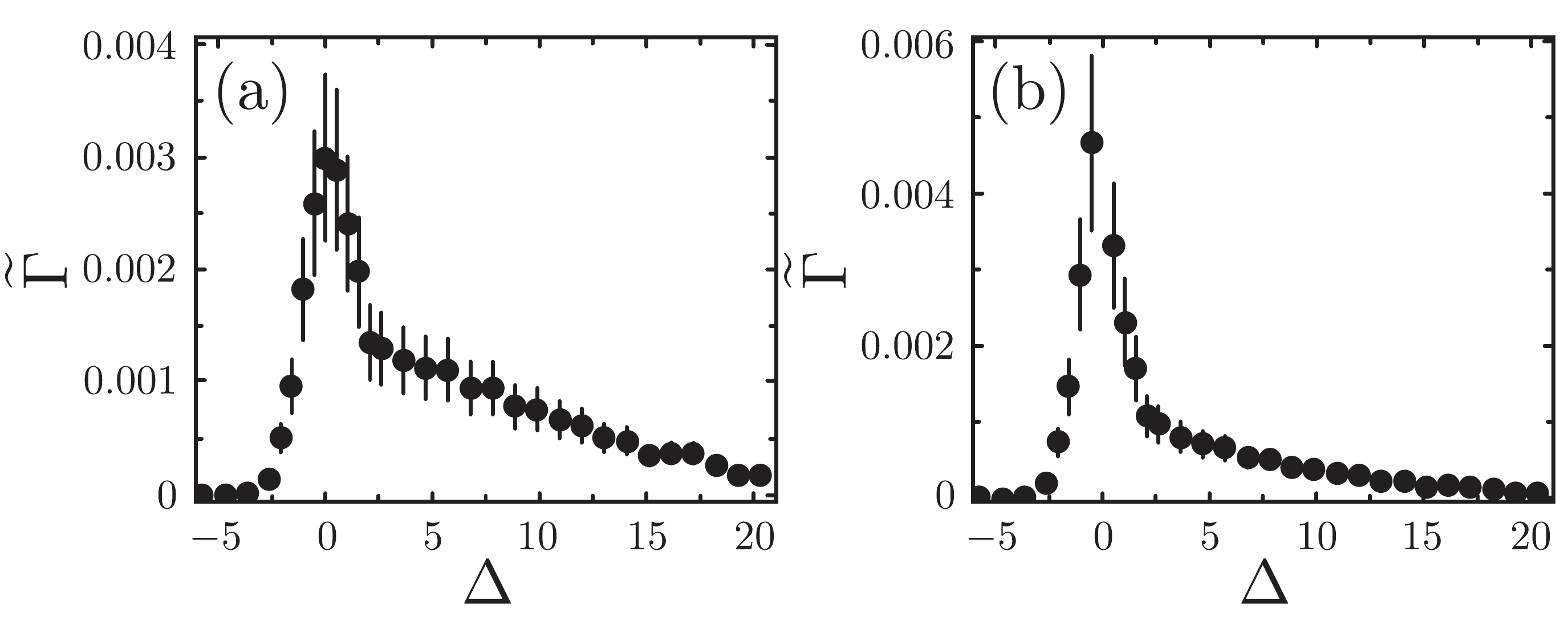}
\caption{Spectroscopy. (a) A typical spectrum taken at $\lnkfa=0.35(5)$ [or $\ln{(k_{\rm{Ff}}a_{\rm{2D}})}=0.00(5)$] after normalization. (b) The same spectrum after rescaling to account for final state interactions using Eq.~(\ref{eq:norm}).
\label{fig:Spectra}}
\end{figure}

\section{RF Spectroscopy}
\paragraph*{Final-state corrections in spectroscopy.}
The high-frequency tail in the spectral density $I_\sigma(\omega_{\rm{rf}})$ of the rf transition to a non-interacting probe state ($\int I_\sigma(\omega_{\rm{rf}})d\omega_{\rm{rf}}=N_\sigma$) is given by \cite{Randeria2010}
\begin{equation}
I_\sigma(\omega_{\rm{rf}})\to\frac{1}{4\pi}\Contact\left( \frac{\hbar}{m} \right)\omega_{\rm{rf}}^{-2}.
\label{eq:spect}
\end{equation}
where $\omega_{\rm{rf}}$ is the detuning from resonance and $\Contact$ is the 2D contact. Following the convention of \cite{Bardon:2014} we introduce the normalized dimensionless detuning $\Delta\equiv\hbar\omega_{\rm{rf}}/E_{\rm{F}}$ and normalized transfer rate $\widetilde{\Gamma}\equiv E_{\rm{F}}I_\sigma/(2\hbar N_\sigma)$ which obeys the sum rule $\int\widetilde{\Gamma}(\Delta)d\Delta=1/2$ where we have identified $I_\sigma=2N_p/(\Omega_R^2\pi t_{\rm{rf}})$. Here $\Omega_R$ is the Rabi frequency and $t_{\rm{rf}}$ is the duration of the rf pulse. Combining this normalization with Eq.~(\ref{eq:spect}) and taking final state interactions into account~\cite{Langmack:2012hr,Frohlich:2012ic}, for $\Delta\gg1$, we have
\begin{equation}
\widetilde{\Gamma}\to\frac{1}{2\pi}\frac{\Contact}{Nk_{\rm{F}}^2}\Delta^{-2}\times\frac{\ln{^2(\widetilde{E}_B/E_B)}}{\ln{^2(\Delta E_{\rm{F}}/\widetilde{E}_B})+\pi^2}
\label{eq:norm}
\end{equation}
where $\widetilde{E}_B$ is the binding energy of the most weakly bound state of atoms in the final state and $E_B$ is the binding energy of the confinement induced dimer. For spectroscopic measurements we use the Fermi energy $E_{\rm{Ff}}$ and wave vector $k_{\rm{Ff}}$ of a fully depolarized sample. We use Eq.~(\ref{eq:norm}) to determine $\Contact(t_h)/(Nk_{\rm{Ff}}^2)$. The time-resolved data in Fig.~4 of the main text is taken at a single frequency $\Delta=9.5$.

Figure \ref{fig:Spectra} shows a representative spectrum at $\lnkfa=0.35(5)$ before and after accounting for final state interactions. We ensure that an appropriate Rabi frequency is chosen such that for $\Delta\gtrsim5$ (where we extract a value of $\Contact$) we probe the transition in the linear regime. For a full spectrum taken within this linear regime, we measure $\int\widetilde{\Gamma}(\Delta)d\Delta=0.50(5)$ for $\lnkfa\approx0.5$ in good agreement with the sum rule.

\paragraph*{Direct comparison of contact and magnetization dynamics.}
In order to compare $\Contact$ and $M_{xy}$ dynamics a spin-refocussing $\pi$ pulse can be applied to the $\Contact$ measurement described in the main text at time $t_h/2$. The evolution of $\Contact$ with a spin-refocussing pulse is a factor $4^{-1/3}$ slower than without and otherwise identical. Comparing to a $1-|M|^\beta$ dependence, we find $\beta=1.4(2)$ between the fully correlated $\beta=1$ and fully uncorrelated $\beta=2$ limits. We use this value of $\beta$ to model the dynamics of $\Contact$.

\bibliography{bibLR2D}